\documentclass[preprint,aps]{revtex4-1}
\pdfoutput=1

\usepackage{amsmath,amssymb,amsfonts,dcolumn,color,graphicx,graphics,latexsym,placeins,epsfig}
\usepackage{epsfig}
\usepackage{bm}
\usepackage{slashed}
\usepackage{latexsym}
\usepackage{natbib}
\usepackage{url}
\usepackage{dcolumn}
\usepackage{color}
\usepackage{amsfonts,amssymb,amsmath}
\usepackage{graphicx,epsfig}
\usepackage{psfrag}
\usepackage{subfigure}
\usepackage{tabularx}
\usepackage{hyperref}
\hypersetup{colorlinks=true}

\newcommand{\be}{\begin{equation}}
\newcommand{\ee}{\end{equation}}
\newcommand{\ba}{\begin{eqnarray}}
\newcommand{\ea}{\end{eqnarray}}

\begin{document}

\title{Pulsar kicks in ultralight dark matter background induced by neutrino oscillation}
\author{Gaetano Lambiase$^{1}$\footnote{lambiase@sa.infn.it}}
\author{Tanmay Kumar Poddar$^{2}$\footnote{tanmay.poddar@tifr.res.in}}
\affiliation{${}^{1}$ \it Dipartimento di Fisica E.R. Caianiello, Universit\`a di Salerno, Via Giovanni Paolo II, 132 I-84084 Fisciano (SA), Italy.\\
${}^{1}$ INFN, Gruppo collegato di Salerno, Italy.}

\affiliation {${}^{2}$ \it Department of Theoretical Physics, Tata Institute of Fundamental Research, Mumbai- 400005, India}


\begin{abstract}
The interaction of neutrinos with ultralight scalar and vector dark matter backgrounds induce a modification of the neutrino dispersion relation. The effects of this modification are reviewed in the framework of asymmetric emission of neutrinos from the supernova core, and, in turn, of pulsar kicks. We consider the neutrino oscillations, focusing in particular to active-sterile conversion. The ultralight dark matter induced neutrino dispersion relation contains a term of the form $\delta {\bf \Omega}\cdot \hat{{\bf{p}}}$, where $\delta {\bf \Omega}$ is related to the ultralight dark matter field and $\hat{{\bf p}}$ is the unit vector along the direction of neutrino momentum. The relative orientation of ${\bf p}$ with respect to $\delta {\bf \Omega}$ affects the mechanism for the generation of the observed pulsar velocities. We obtain the resonance condition for the active-sterile neutrino oscillation in ultralight dark matter background and calculate the star parameters in the resonance surface so that both ultralight scalar and vector dark matter backgrounds can explain the observed pulsar kicks. The asymmetric emission of neutrinos in presence of ultralight dark matter background results gravitational memory signal which can be probed from the future gravitational wave detectors such as adLIGO (advanced LIGO), adVIRGO (advanced VIRGO), DECIGO (DECi-hertz Interferometer Gravitational wave Observatory), BBO (Big Bang Observer), and ET (Einstein Telescope). We also establish a relation between the ultralight dark matter parameters and the Lorentz and CPT invariance violation parameters.
\end{abstract}

\maketitle
\section{Introduction}
\label{sec1}
The ultralight scalar and vector particles are among the favourite Dark Matter (DM) candidates as they can solve some of the small scale structure problems of the universe and evade the direct detection constraints on DM \cite{Hu:2000ke,Marsh:2015wka,Cembranos:2016ugq,DelPopolo:2016emo,Robles:2018fur,Chadha-Day:2021szb,Amin:2022pzv}. Being ultralight, they can have a de Broglie wavelength of the order of galactic scale. If we consider the local DM density as $\rho_{\rm{DM}}\sim 0.4~\rm{GeV}/\rm{cm^3}$, then the number density of DM is $n_{\rm{DM}}\sim 10^{30}/\rm{cm^3}$ for DM mass $\sim 10^{-22}~\rm{eV}$. Such a large value of $n_{\rm{DM}}$ within the de Broglie wavelength implies that they may oscillate coherently as a single classical field. These classical fields can oscillate in a cosmological scale and at late time its energy density redshifts like a cold DM. The Ultralight Scalar Dark Matter (USDM) and Ultralight Vector Dark Matter (UVDM) can have only gravitational interaction with the visible sector or they can have very small coupling with the Standard Model (SM) particles. There are many ongoing and upcoming experiments which are trying to search for these particles. They can also be probed from the variation of fundamental constants in nature \cite{Budker:2013hfa,Stadnik:2014tta,Stadnik:2015kia}. The ultralight DM can be probed from the Time Of Arrival (TOA) of pulses coming from rapidly rotating pulsars \cite{Poddar:2020qft,Kaplan:2022lmz}. Gravitational Waves (GWs) observed by LIGO/Virgo put bounds on ultralight DM as well \cite{Kopp:2018jom,Poddar:2023pfj}. Several tests of gravity in the solar system scale obtain interesting constraints on UVDM and USDM \cite{KumarPoddar:2019jxe,KumarPoddar:2020kdz,KumarPoddar:2019ceq,Poddar:2021sbc,Poddar:2021ose,Fedderke:2021kuy,Poddar:2023bgk,Tsai:2021irw,Tsai:2021lly,Fedderke:2022ptm}. They can also be probed from the neutrino oscillation experiments \cite{Alonso-Alvarez:2021pgy,Alonso-Alvarez:2023tii,Gherghetta:2023myo}. The ultralight (pseudo) scalar particles such as axion, majoron, dilaton and the ultralight vector particles such as $L_i-L_j$ vector gauge boson, dark photon are the potential DM candidates.

There are several astrophysical and cosmological observations which can put constraints on Ultra-Light DM (ULDM). Existing measurements of the sizes and radial velocities of ultra faint dwarf galaxies Seague 1 and Seague 2 put bounds on the ULDM as $m_{\mathrm{ULDM}}>3\times 10^{-19}~\mathrm{eV}$ \cite{Dalal:2022rmp}. The ULDM affects the small scale matter power spectrum and consequently the angular scale of the CMB acoustic peaks and anisotropies are affected. Measurements of Planck 2015 and WiggleZ put bounds on ULDM mass as $m_{\mathrm{ULDM}}\gtrsim 10^{-24}~\mathrm{eV}$ \cite{Hlozek:2014lca,Hlozek:2017zzf}. Lyman-$\alpha$ forest also puts bound on the mass of ULDM as $m_{\mathrm{ULDM}}>(2.0-20)\times 10^{-21}~\mathrm{eV}$ \cite{Nori:2018pka,Rogers:2020ltq}. Measurements of the shear correlation of galaxies from DES-Y1 and Planck data put bounds on the ULDM mass as $m_{\mathrm{ULDM}}>10^{-23}~\mathrm{eV}$ \cite{Dentler:2021zij}. The non detection of black hole superradiance limits the ULDM mass as $7\times 10^{-20}~\mathrm{eV}<m_{\mathrm{ULDM}}<10^{-16}~\mathrm{eV}$ for supermassive black holes \cite{Arvanitaki:2010sy,Baryakhtar:2017ngi} and $7\times 10^{-14}~\mathrm{eV}<m_{\mathrm{ULDM}}<2\times 10^{-11}~\mathrm{eV}$ for stellar mass black holes \cite{Brito:2017zvb}. Bounds from other observations and experiments are mentioned in \cite{Banares-Hernandez:2023axy} and the references therein. It is important to note that these constraints rely on certain simplifications in their methodology and assumptions. 

The USDM can be produced from the standard misalignment mechanism \cite{Hui:2016ltb}. The origin of the mass of the UVDM particles is model dependent, where the mass can be either produced from the Higgs mechanism or Stueckelberg mechanism \cite{Nelson:2011sf}. They can also be produced from the clockwork mechanism \cite{Joshipura:2020ibd} or from the FI (Fayet-Ilioupoulous) term in supersymmetry theory \cite{Fayet:1974jb,Fayet:2017pdp}. The UVDM can be produced through misalignment mechanism \cite{Alonso-Alvarez:2019ixv}, freeze-in mechanism \cite{Redondo:2008ec} and fluctuations during inflation \cite{Graham:2015rva}. The misalignment mechanism was first introduced for axion production. Unlike axions, in the $H\gg m_{\rm{VDM}}$ limit, the energy density of UVDM redshifts as $1/a^2$, where $a$ is the scale factor. Thus the initial energy density of UVDM dilutes during inflation and the misalignment mechanism badly fails in producing UVDM. This crisis can be avoided by choosing $\mathcal{O}(1)$ nonminimal coupling to gravity. This makes the effective Lagrangian conformally invariant that remains unaffected by the expansion of the universe.

In this paper, we investigate the effects of the presence of UVDM and USDM background on pulsar kicks induced by neutrino oscillations in matter. Observations show that pulsars have a very high velocity that varies from an average of $450~\rm{km/s}$ up to values greater than $1000~\rm{km}/s$ compared with the surrounding stars whose velocities are roughly $30~\rm{km/s}$ \cite{Lyne:1994az}. This suggests that nascent pulsars undergo to a kind of kick during their birth. The physical description underlying this process is that during the cooling of a proto-neutron star, neutrinos carry away a large percentage, about $99\%$ of the gravitational binding energy, which is of the order $3\times 10^{53}~\rm{erg}$ while the momentum carried by the neutrinos is $10^{43}~\rm{g~cm/s}$. A fractional anisotropy of $~1\%$ of the outgoing neutrino momenta distribution is therefore suffice to account for the pulsar kick. 

A mechanism to generate the pulsar velocity involving active neutrino oscillations has been proposed in \cite{Kusenko:1996sr,Kusenko:1998bk,Barkovich:2002wh}. Here, it was shown that, under suitable conditions, it is possible that the resonant oscillation $\nu_e\rightarrow \nu_{\mu,\tau}$ may occur in the region between the $\nu_e$ neutrinosphere and $\nu_{\mu,\tau}$ neutrinosphere. Owing to neutral and charged current interactions, electron neutrinos $\nu_e$ are trapped by the medium, but muon and tau neutrinos $\nu_{\mu,\tau}$ generate via oscillations can escape from the proto-star being outside of their neutrinosphere. The resonance surface acts, in such a case, as an effective spherical $\nu_{\mu,\tau}$ neutrinosphere. Such a resonance surface, however is distorted if, for example, a magnetic field $\textbf{B}$ is present, thus the energy flux is generated anisotropically. More specifically, the term  $\textbf{B}.~\textbf{p}$ induced by a magnetized medium modifies the resonance condition in matter, where $\textbf{p}$ denotes the neutrino momentum. Hence, the relative orientation of the neutrino momentum with respect to $\textbf{B}$ giving rise to the asymmetry in the neutrino emission. However, in this mechanism one needs a very heavy tau neutrino mass $(\sim 100~\rm{eV})$ to explain the pulsar kick. This value of neutrino mass is in strong tension with the bounds obtained from neutrino oscillation experiments and cosmology \cite{Planck:2018vyg,ParticleDataGroup:2020ssz}. This bound can be evaded if one considers the active-sterile neutrino oscillation and is discussed in \cite{Barkovich:2004jp} which requires a large magnetic field $(10^{17}~\rm{G}-10^{18}~\rm{G})$ for pulsar kicks to happen. This phenomenon can also be explained from the violation of CPT and Lorentz invariance \cite{Lambiase:2005iz}, neutrino spin flavour oscillation in gravitational fields \cite{Lambiase:2004qk} and violation of equivalence principle \cite{Horvat:1998st} (see also \cite{Qian:1997sw,Akhmedov:1997qb,Grasso:1998tt,Janka:1998kb,Farzan:2005yp,Loveridge:2003fy,Katsuda:2017hco,Holland-Ashford:2017tqz,Fragione:2023mpe}) .

In this paper, we consider the active-sterile neutrino oscillation to explain the pulsar kick and the asymmetry in the neutrino emission from the proto-neutron star is generated due to the ultralight DM background. In fact the relative orientation of neutrino momenta with respect to the direction of ultralight DM field induces an asymmetric emission of the neutrinos, contributing in such a way to the pulsar kick. This is in analogy with the model of asymmetric emission of neutrinos induced by proto-neutron star's magnetic field $\textbf{B}$. It must be pointed out that hydrodynamical asymmetries in the supernova collapse may be able to generate such large velocities \cite{Nordhaus:2010ub,Scheck:2003rw,Scheck:2006rw} (see also \cite{Lai:2000pk,Wex:1999kr,Colpi:2002cu,Johnston:2005ka,Wang:2005jg,Ng:2007aw,Ng:2007te}). Recently, it has been studied that the magnetic rocket effect where an off-centre dipole or an asymmetric field can also produce the pulsar kick \cite{Agalianou:2023lvv}. The proto-neutron star can also be accelerated by asymmetric neutrino emission based on detail multi dimensional core-collapse supernovae model with full Boltzmann neutrino transport \cite{Nagakura:2019evv}.

The paper is organized as follows. In Section \ref{sec2}, we discuss the pulsar kicks in presence of UVDM background induced by neutrino oscillation. In Section \ref{sec3}, we similarly explain the pulsar kick in presence of USDM background. The detection of gravitational memory signal in GW detectors due to the asymmetric neutrino emission is discussed in Section \ref{extra}. We also establish a connection between the ultralight DM parameters and Standard Model Extension (SME) parameters. Finally we conclude our results in Section \ref{sec4}.

We use natural system of units throughout the paper, where $\hbar=c=1$.
\section{Pulsar kicks induced by active-sterile neutrino oscillation in presence of ultralight vector dark matter background}\label{sec2}
The Ultralight Vector Dark Matter (UVDM) can interact with the neutrino as described by the Lagrangian 
\begin{equation}
-\mathcal{L}_{\rm{int}}=g^\prime_{\alpha\beta}\bar{\nu}_\alpha\gamma^\mu\nu_\beta A^\prime_\mu, \hspace{0.5cm} \alpha,\beta=e,\mu,\tau,
\label{eq:2}
\end{equation}
with $g^\prime_{\alpha\beta}(=g^{\prime*}_{\beta\alpha})$ is a dimensionless coupling strength in the neutrino flavour basis, $\nu_{\alpha(\beta)}$ is the neutrino flavour eigenstate and $A^\prime_\mu$ denotes the UVDM field. The neutrino energy dispersion relation changes due to the UVDM-neutrino interaction as 
\begin{equation}
E_\nu=\Big((\textbf{p}+g^\prime_{\alpha\beta}\textbf{A}^\prime)^2+m^2\Big)^\frac{1}{2}\simeq |\textbf{p}|+g^\prime_{\alpha\beta}\hat{\textbf{p}}\cdot \textbf{A}^\prime+\frac{m^2}{2|\textbf{p}|}+\mathcal{O}({g^\prime}^2_{\alpha\beta}{A^\prime}^2),
\end{equation}
where we set $\left\langle A^\prime_0\right \rangle=0$ using the gauge freedom. Here, $\hat{\textbf{p}}=\textbf{p}/p$ with $p=|\textbf{p}|$ is the direction of the neutrino flux and $m$ denotes the neutrino mass. The UVDM may oscillate coherently as a single classical field, given by
\begin{equation}
|\textbf{A}^\prime(t)|=\frac{\sqrt{2\rho_{\rm{DM}}}}{m_{A^\prime}}\cos(m_{A^\prime}t+\theta),
\end{equation}
where $m_{A^\prime}$ is the mass of the UVDM, $\rho_{\rm{DM}}=0.4~\rm{GeV}/cm^3$ is the local DM density and $\theta$ is the initial phase of the oscillatory DM field.
 Therefore, the effective flavour potential induced by the UVDM background becomes 
\begin{equation}
\xi_{\alpha\beta} = g^\prime_{\alpha\beta}\hat{\textbf{p}}\cdot \textbf{A}^\prime=g^\prime_{\alpha\beta}\frac{\sqrt{2\rho_{\rm{DM}}}}{m_{A^\prime}}\hat{\textbf{p}}\cdot\hat{\mathbf{\Omega}}=\mathbf{\Omega}\cdot \hat{\textbf{p}},
\label{eq:4}
\end{equation}
where we consider that the mass of UVDM is smaller than the inverse time scale of DM field, $\mathbf{\Omega}=g^\prime_{\alpha\beta}\frac{\sqrt{2\rho_{\rm{DM}}}}{m_{A^\prime}}\hat{\mathbf{\Omega}}$, and $\hat{\mathbf{\Omega}}$ denotes the direction of $\textbf{A}^\prime$.

We consider the time/length scales of ULDM variation (coherent DM oscillations) very large compared to the supernova collapse process. Therefore, the ULDM behaves as a long range field. Under this consideration, we can write the mass of DM as $m_{\mathrm{ULDM}}\lesssim\frac{1}{T}$, where $T$ is the time taken for the DM to traverse the distance $(R\sim 10~\mathrm{km})$ of the proto-star. In this limit we can omit the time oscillation of the DM field. Hence, the above relation becomes $m_{\mathrm{ULDM}}\lesssim \frac{v}{R}$, where $v\sim 10^{-3}$ is the typical velocity of the DM. Putting the values of $v$ and $R$, we obtain $m_{\mathrm{ULDM}}\lesssim 1.98\times 10^{-14}~\mathrm{eV}$. Hence, the oscillation period of the DM becomes $T_{\mathrm{osc}}=\frac{2\pi}{m_{\mathrm{ULDM}}}\gtrsim0.2~\mathrm{s}$. So our results are valid for the mass of ULDM $m_{\mathrm{ULDM}}\lesssim 1.98\times 10^{-14}~\mathrm{eV}$.

Hence, the effective Hamiltonian describing the neutrino evolution reads
\begin{equation} \label{Heff}
H =  \frac{1}{2E^{}_{\nu}} U \left(\begin{matrix} 0 & 0 & 0 \\ 0 & \Delta m^2_{21} & 0 \\ 0 & 0 & \Delta m^2_{31} \end{matrix}\right)  U^\dagger
+\left(\begin{matrix} V_{\nu_e}+\xi_{ee} & \xi_{e\mu}& \xi_{e\tau} \\ \xi^{*}_{e\mu} & V_{\nu_\mu}+\xi_{\mu\mu} & \xi_{\mu\tau} \\ \xi^{*}_{e\tau} & \xi^{*}_{\mu\tau} & V_{\nu_\tau}+\xi_{\tau\tau} \end{matrix}\right)\,,
\end{equation}
where $U$ is for the flavour mixing matrix of the three generation of neutrinos in vacuum, $\Delta m^2_{21}$ and $\Delta m^2_{31}$ are the neutrino mass squared differences in vacuum, $V_{\nu_e,\nu_\mu,\nu_\tau}$ is the usual matter potential which arises due to the interaction of neutrinos with ordinary matter and $\xi_{\alpha\beta}$ is the potential arises due to the interaction of neutrino with UVDM background.

As discussed in the Section \ref{sec1}, the explanation of pulsar velocity by means of active neutrino oscillation requires that the resonant conversion $\nu_e\leftrightarrow \nu_{\mu,\tau}$ occurs between two different neutrinospheres. Such an approach, however, leads to neutrino masses that are not compatible with the limits on the masses of standard electroweak neutrinos. These limits can be avoided by considering sterile neutrinos, that may have only a small mixing angle with the ordinary neutrinos. In what follows we shall therefore confine to the active-sterile conversion of neutrinos. 

The energy flux $\textbf{F}_S(\varphi)$ emitted by nascent star allows to estimate the anisotropy of the outgoing neutrinos. Following \cite{Barkovich:2002wh}, we can write the asymmetry of the neutrino momentum as
\begin{equation}
\frac{\Delta p}{p}=\frac{1}{6}\frac{\int^\pi_0 \textbf{F}_S(\varphi)\cdot\delta\hat{\mathbf{\Omega}}dS}{\int^\pi_0 \textbf{F}_S(\varphi)\cdot\hat{\mathbf{n}}dS}\simeq -\frac{1}{18}\frac{\varrho}{r_{\rm{res}}}
\label{eq:5}
\end{equation}
where the $1/6$ factor is introduced because only one flavour of neutrinos is responsible for the anisotropy, and therefore it carries out only $1/6$ of the total energy. Here, $dS$ denotes the elemental surface and $\hat{\mathbf{n}}$ is the unit vector orthogonal to $dS$. Also, $\delta\hat{\mathbf{\Omega}}=\frac{\mathbf{\Omega}_f-\mathbf{\Omega}_s}{|\mathbf{\Omega}_f-\mathbf{\Omega}_s|}$ and $\varrho$ is the parameter that accounts for the radial deformation of the effective surface of resonance generated by the UVDM background. The latter shifts the resonance point $r_{\rm{res}}$ to $r(\varphi)=r_{\rm{res}}+\varrho\cos\varphi$, where $\cos\varphi=\delta\hat{\mathbf{\Omega}}\cdot\hat{\mathbf{p}}$ and $\varrho \ll r_{\rm{res}}$ (for the second equality in Eq. \ref{eq:5}, see Appendix \ref{app1}).

We shall confine ourselves to the conversion $\nu_f\leftrightarrow \nu_s$, $f=e,\mu,\tau$ and the neutrino evolution equation becomes
 \begin{equation}\label{eq:6}
i\frac{d}{d r}\left(\begin{array}{c}
                           \nu_{f} \\
                              \nu_s \end{array}\right)={\cal H}\left(\begin{array}{c}
                           \nu_{f} \\
                             \nu_s\end{array}\right)\,,
 \end{equation}
where the effective Hamiltonian ${\cal H}$ is defined as
\begin{equation}\label{eq:7}
{\cal H}=\left[\begin{array}{cc}
 V_{\nu_f}-c_2\delta + \xi_{ff} & s_2\delta \vspace{0.05in} \\
 s_2\delta & c_2\delta
\end{array}\right]\,,
\end{equation}
up to terms proportional to identity matrix. Also, we assume that the coupling of sterile neutrino with UVDM is smaller than the coupling of active neutrino with UVDM and hence, we can write $\xi_{ff}-\xi_{ss}\approx \xi_{ff}$. Therefore, the resonance condition becomes
\begin{equation}
2\delta c_2=V_{\nu_f}+\xi_{ff},
\label{eq:8}
\end{equation}
where $\delta=\frac{\Delta m^2}{4p}$, $c_2=\cos2\theta$, $s_2=\sin2\theta$, and the potential $V_{\nu_f}$ is given by \cite{Caldwell:1999zk}
\begin{eqnarray}\label{eq:9}
 V_{\nu_e} &=& -V_{{\bar \nu}_e}=V_0(3Y_e-1+4Y_{\nu_e})\,, \\
 V_{\nu_{\mu,\tau}} &=& -V_{{\bar
 \nu}_{\mu,\tau}}=V_0(Y_e-1+2Y_{\nu_e})\,, \label{eq:10} \\
 V_{\nu_s} &=& 0\,,
\end{eqnarray}
where $Y_e$  and $Y_{\nu_e}$ denote the electron and electron neutrino fractions respectively and we assume that the muon and tau neutrino fractions are smaller compared to $Y_{\nu_e}$. Here $V_0$ is defined as
\begin{equation}\label{eq:11}
  V_0=\frac{G_F\rho}{\sqrt{2}m_n}=\,
  3.8\,\mbox{eV}\Big(\frac{\rho}{10^{14}\mbox{g/cm}^3}\Big)\,,
\end{equation}
where $m_n$ is the nucleon mass, $G_F$ is the Fermi constant and $\rho$ is the matter density (for typical values of $\rho\sim (10^{11}-10^{14})~\rm{g}/{cm^3}$ in the proto-star). Since, $\xi_{ff}=\delta\hat{\mathbf{\Omega}}\cdot\hat{\textbf{p}}$ changes sign under the transformation $\textbf{p}\rightarrow -\textbf{p}$, it may deform the resonance surface, which is the necessary condition for the asymmetry of the outgoing neutrino momenta to occur.

To evaluate $\varrho$, we expand Eq. \ref{eq:8} about $r=r_{\rm{res}}+\varrho\cos\varphi$. We also expand $p$ and $V_{\nu_f}$ about their resonance values as $p=p_{\rm{res}}+\delta_p$ and $V_{\nu_f}=V_{\nu_{f\rm{res}}}+\delta_{V_{\nu_f}}$, where 
\begin{equation}
\delta_p=\frac{dp}{dr}\bigg|_{r_{\rm{res}}}\delta r_{\rm{res}}=\frac{d\ln p}{dr}\bigg|_{r_{\rm{res}}}p_{\rm{res}}\delta r_{\rm{res}}=h^{-1}_p p_{\rm{res}}\varrho\cos\varphi,
\label{eq:12}
\end{equation}
and 
\begin{equation}
\delta_{V_{\nu_f}}=\frac{dV_{\nu_f}}{dr}\bigg|_{r_{\rm{res}}}\delta r_{res}=\frac{d\ln V_{\nu_f}}{dr}\bigg|_{r_{\rm{res}}}V_{\nu_{f\rm{res}}}\delta r_{\rm{res}}=h^{-1}_{V_{\nu_f}} p_{\rm{res}}\varrho\cos\varphi,
\label{eq:13}
\end{equation}
where $h_p^{-1}=\frac{d\ln p}{dr}\big|_{r_{\rm{res}}}$ and $h^{-1}_{V_{\nu_f}}=\frac{d\ln V_{\nu_f}}{dr}\big|_{r_{\rm{res}}}$.

Using the resonance condition in absence of UVDM background, we obtain the expression of $\varrho$ from Eq. \ref{eq:8} as 
\begin{equation}
\varrho=-\frac{g^\prime_{ff}\frac{\sqrt{2\rho_{\rm{DM}}}}{m_{A^\prime}}}{V_{\nu_f}(h^{-1}_p+h^{-1}_{V_{\nu_f}})},
\label{eq:14}
\end{equation}
where all the quantities are measured at the resonance point. Hence, from Eq. \ref{eq:5} we can write the neutrino momentum asymmetry as 
\begin{equation}
\frac{\Delta p}{p}=\frac{1}{18}\times\frac{g^\prime_{ff}\frac{\sqrt{2\rho_{\rm{DM}}}}{m_{A^\prime}}}{V_{\nu_f}r_{\rm{res}}(h^{-1}_p+h^{-1}_{V_{\nu_f}})}.
\label{eq:15}
\end{equation}
The fractional asymmetry $\Delta p/p\sim 1\%$ implies 
\begin{equation}
g^\prime_{ff}\frac{\sqrt{2\rho_{\rm{DM}}}}{m_{A^\prime}}=0.18 V_{\nu_f}r_{\rm{res}}\Big(\frac{1}{h_p}+\frac{1}{h_{V_{\nu_f}}}\Big).
\label{eq:16}
\end{equation}
The quantities $h_p$ and $h_{V_{\nu_f}}$ are calculated for a specific model of the proto-star. We assume that the inner core of a proto-star is described by a polytropic gas of relativistic nucleons, with equation of state $P=K\rho^\Gamma$, where $P$ is the pressure, $\rho$ is the matter density, $\Gamma=4/3$ is the adiabatic index and $K$ is a constant given by $K=T_c/m_n\rho_c^{1/3}\simeq 5.6\times 10^{-5}~\rm{MeV}^{-4/3}$ with $T_c=40~\rm{MeV}$ and $\rho_c=10^{14}~\rm{g/cm^3}$ are the temperature and the matter density of the core respectively. By defining $x=\frac{r_c}{r}$, $\alpha=(1-\mu)\lambda_\Gamma$, $\beta=(2\mu-1)\lambda_\Gamma$, and $\gamma=1-\mu\lambda_\Gamma$ we can write (see Appendix \ref{app2}) \cite{Barkovich:2002wh}
\begin{equation}
\rho^{\Gamma-1}(x)=\rho_c^{\Gamma-1}(\alpha x^2+\beta x+\gamma),
\label{eq:17}
\end{equation}
where 
\begin{equation}
\lambda_\Gamma=\frac{GM_c}{r_c\rho_c^{\Gamma-1}}\frac{\Gamma-1}{K\Gamma}\simeq 0.29\frac{2GM_c}{\rm{km}}\frac{10~\rm{km}}{r_c}\frac{40~\rm{MeV}}{T_c},
\label{eq:18}
\end{equation}
where $G$ is the Newton's gravitation constant, $r_c=10~\rm{km}$ is the core radius, $M_c\simeq M_\odot$ is the mass of the core ($M_\odot$ is the solar mass) and we use $\Gamma=4/3$. The parameter $\mu$ is determined by setting $\rho(R_s)=0$ (here $R_s$ denotes the radius of the star) and we get
\begin{equation}
\mu=\Big[\frac{R_s}{\lambda_\Gamma(R_s-r_c)}-\frac{r_c}{R_s}\Big]\frac{R_s}{R_s-r_c}.
\label{eq:19}
\end{equation}
The temperature profile $T(r)$ and matter density $\rho(r)$ are related as \cite{Barkovich:2002wh} (see Appendix \ref{app2})
\begin{equation}
\frac{dT^2}{dr}=-\frac{9\kappa L_c}{\pi r^2}\rho,
\label{eq:20}
\end{equation}
where $L_c\sim 9.5\times 10^{51}~\rm{erg}/s$ is the core luminosity, and $\kappa=5.6\times 10^{-9}~\rm{cm^4/erg^3~s^2}$. The solution of Eq. \ref{eq:20} is
\begin{equation}
T(r)=T_c\sqrt{2\lambda_c\Big[\chi(r_c/r)-\chi(1)+1\Big]},
\label{eq:21}
\end{equation}
where $\chi(x)$ is a polynomial function of $x$ and is given as \cite{Barkovich:2002wh}
\begin{equation}
\chi(x)=\gamma^3 x+\frac{3}{2}\beta\gamma^2x^2+\gamma(\alpha\gamma+\beta^2)x^3+\frac{\beta}{4}(6\alpha\gamma+\beta^2)x^4+\frac{3\alpha}{5}(\alpha\gamma+\beta^2)x^5+\frac{\beta\alpha^2}{2}x^6+\frac{\alpha^3}{7}x^7,
\label{eq:22}
\end{equation}
where $x=r_c/r$ as defined earlier and 
\begin{equation}
\lambda_c=\frac{9}{2\pi}\frac{\kappa L_c\rho_c}{T^2_cr_c}\sim 1.95\frac{\rho_c}{10^{14}~\rm{g/cm^3}}\frac{10~\rm{km}}{r_c}\Big(\frac{40~\rm{MeV}}{T_c}\Big)^2.
\label{eq:23}
\end{equation}
At $r=r_c$ or $x=1$, we can calculate $T(r_c)$ by solving Eq. \ref{eq:20}, using Eq. \ref{eq:23}. We obtain $T(r_c)=T_c\sqrt{2\lambda_c}$ which is consistent with Eq. \ref{eq:21}.

Assuming the neutrinos are in thermal equilibrium with the medium, so that the neutrino average energy is proportional to the temperature at the emission point, i.e; $p\sim T$ and $V_{\nu_f}\sim \rho$. Therefore, the inverse characteristic lengths $h_p^{-1}$ and $h_{V_{\nu_f}}^{-1}$ can be written as $h^{-1}_p\equiv h^{-1}_T$ and $h^{-1}_{V_{\nu_f}}\equiv h^{-1}_\rho$. Hence, at the resonance we can write
\begin{equation}
h^{-1}_T=\frac{d\ln T}{dr}=-2\lambda_c\frac{\rho(r_{\rm{res}})}{\rho_c}\Big(\frac{T_c}{T(r_{\rm{res}})}\Big)^2\frac{x_{\rm{res}}}{r_{\rm{res}}},
\label{eq:24}
\end{equation}
\begin{equation}
h^{-1}_\rho=\frac{d\ln \rho}{dr}=-3\Big(\frac{\rho_c}{\rho(r_{\rm{res}})}\Big)^{1/3}(2ax_{\rm{res}}+b)\frac{x_{\rm{res}}}{r_{\rm{res}}}.
\label{eq:25}
\end{equation}
The UVDM background is relevant on pulsar kicks for resonances occurring at
\begin{equation}
\rho(r_{\rm{res}})=\rho_c.
\label{eq:26}
\end{equation}
Adding Eq. \ref{eq:24} and Eq. \ref{eq:25} and using Eq. \ref{eq:26}, we obtain 
\begin{equation}
h^{-1}_T+h^{-1}_\rho=-\frac{x_{\rm{res}}}{r_{\rm{res}}}\lambda_\Gamma\eta,
\label{eq:27} 
\end{equation}
where 
\begin{equation}
\eta=\frac{2\lambda_c}{\lambda_\Gamma}\epsilon^2+3(2\mu-1)-6(\mu-1)x_{\rm{res}},
\label{eq:28} 
\end{equation}
with $\epsilon=T_c/T(r_{\rm{res}})$. Therefore, we can write Eq. \ref{eq:16} as
\begin{equation}
\Big|g^\prime_{ff}\frac{\sqrt{2\rho_{\rm{DM}}}}{m_{A^\prime}}\Big|=0.18V_{\nu_f}x_{\rm{res}}\lambda_\Gamma\eta.
\label{eq:29} 
\end{equation}
At resonance, we can write Eq. \ref{eq:17} as $\alpha x^2_{\rm{res}}+\beta x_{\rm{res}}+\gamma=1$. The solution of this equation becomes $x_{\rm{res}}=1,\mu/(\mu-1)$. Hence, we obtain Eq. \ref{eq:28} for these two solutions of $x_{\rm{res}}$ as
\begin{equation}
\epsilon=\Big[(\eta-3)\frac{\lambda_\Gamma}{2\lambda_c}\Big]^{1/2},\hspace{0.2cm} \textrm{for}~ x_{\rm{res}}=1 \hspace{0.2cm} \&~ \epsilon=\Big[(\eta+3)\frac{\lambda_\Gamma}{2\lambda_c}\Big]^{1/2},\hspace{0.2cm} \textrm{for}~ x_{\rm{res}}=\frac{\mu}{\mu-1} \hspace{0.2cm}. 
\label{eq:30} 
\end{equation}
At this stage, we need to fix $V_{\nu_f}$. We consider the active-sterile antineutrino oscillation in the $2-4$ plane so that the corresponding parameters are somewhat less stringent from other experiments. For $Y_e\sim Y_\nu\sim \mathcal{O}(0.1)$, $V_{\nu_\mu}\sim 0.7~V_0$. Therefore, at resonance Eq. \ref{eq:29} becomes
\begin{equation}
\eta=\frac{6.40\times 10^{-12}}{x_{\rm{res}}},
\label{eq:31} 
\end{equation}
where we use $g^\prime_{ff}\sim 3.63\times 10^{-24}$, $m_A^\prime\sim 1.02\times 10^{-14}~\mathrm{eV}$ \cite{Alonso-Alvarez:2023tii}. For $x_{\rm{res}}=1$, $\eta=6.40\times 10^{-12}$ and $\epsilon$ does not admit a real solution. However, for $\mu>1$, $x_{\rm{res}}\sim \mathcal{O}(1)$, $\eta\simeq 6.40\times 10^{-12}$, and hence, $\epsilon\simeq 0.47$ is real. The parameter $\mu$ can be determined from Eq. \ref{eq:21} as follows. At resonance we can write Eq. \ref{eq:21} as
\begin{equation}
\Big(\frac{T(r_{\mathrm{res}})}{T_c}\Big)^2=\frac{1}{\epsilon^2}=2\lambda_c\Big[\chi\Big(\frac{r_c}{r_{\mathrm{res}}}\Big)-\chi(1)+1\Big].
\label{eq:jkl1}
\end{equation}
For real solution of $\epsilon$, we can write Eq. \ref{eq:jkl1} as
\begin{equation}
\frac{1}{(\eta+3)\lambda_\Gamma}=\chi(x_{\mathrm{res}})-\chi(1)+1.
\end{equation} 
Therefore, we can write
\begin{equation}
F(\mu)\equiv\chi\Big(\frac{\mu}{\mu-1}\Big)-\chi(1)+1-\frac{1}{(\eta+3)\lambda_\Gamma}=0,
\label{eq:32} 
\end{equation}
where we put $x_\mathrm{res}=\frac{\mu}{\mu-1}$. As already stated, $\mu$ is a parameter which can be determined from Eq. \ref{eq:19}. The radius of the star $R_s$ is the surface where the density becomes zero i.e; $\rho(R_s)=0$. This condition determines the parameter, $\mu$. The value of $\mu$ is important to obtain the physical solutions of the star parameters where the resonance happens.
\begin{figure}
  \centering
\includegraphics[scale=0.6]{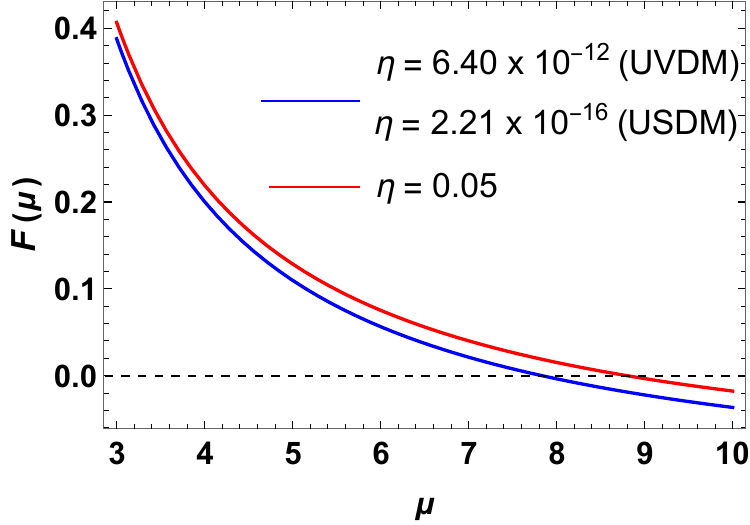}
  \caption{Plot of $F(\mu)$ vs the parameter $\mu$}\label{Fig1}
\end{figure}

In FIG. \ref{Fig1} we obtain the variation of $F(\mu)$ vs. $\mu$ with varying $\eta$ in the range $\eta\in [2.21\times 10^{-16}, 0.05]$. There is a small change in the value of $\mu$ within the range of given $\eta$. The point where the black dashed line cuts the blue solid line is the solution of $\mu$. 

The solutions of Eq. \ref{eq:32} are $\mu\simeq 0.96, 7.85$. The solution $\mu\simeq 0.96$ is discarded as it yields non physical values of star parameters. The physical solution $\mu\simeq 7.85$ yields $T(r_{\rm{res}})\sim 2.13~T_c$, $r_{\rm{res}}\sim 0.87~r_c$ and $R_s\sim 3.07~r_c$.  Using Eq. \ref{eq:8}, we can write at resonance $(\Delta m^2/2p_{\mathrm{res}})\cos2\theta={V_{\nu_f}}_\mathrm{res}+{\xi_{ff}}_\mathrm{res}\approx 2.66~\mathrm{eV}$. Now $p_{\mathrm{res}}\sim T(r_{\mathrm{res}})\sim 2.13~T_c$ and put $T_c=40~\mathrm{MeV}$, we obtain $m_{\nu_s}\sim 21.29~\mathrm{keV}$. The neutrinos are in thermal equilibrium with the medium and the average energy of the neutrino is proportional to the temperature at the emission point. Here we consider the active-sterile mixing angle very small $(\theta\sim 10^{-9})$ so that it can avoid the x-ray bounds , cosmological constraints and contributes negligibly to the DM relic density \cite{Abazajian:2012ys}. The above results are valid for the mass of ULDM $m_{\mathrm{ULDM}}\lesssim 1.98\times 10^{-14}~\mathrm{eV}$. 

We choose the polytropic equation of state as Eq. \ref{eq:17}. For $x\sim \mathcal{O}(1)$, or $r\sim r_c$, $\alpha+\beta+\gamma\sim 1$ and hence from Eq. \ref{eq:17} we obtain $\rho\sim \rho_c$. We consider that the resonance occurs at a surface $(r_{\mathrm{res}})$ near the core radius. Therefore, $\rho({r_\mathrm{res}})=\rho_c$. We obtain various values of star parameters for this consideration as mentioned above. If we were to select a resonance point located at a greater distance from the core radius, our calculations would yield different values for the star's parameters compared to those presented in this paper.

\section{Pulsar kicks induced by active-sterile neutrino oscillation in presence of ultralight scalar dark matter background}\label{sec3}
The Ultralight Scalar Dark Matter (USDM) can couple with neutrino through a derivative coupling as 
\begin{equation}
-\mathcal{L}_{\rm{int}}=g^\prime_{\alpha\beta}\partial_\mu\phi\bar{\nu}_\alpha\gamma^\mu\nu_\beta, \hspace{0.5cm} \alpha,\beta=e,\mu,\tau,
\label{eq:s1}
\end{equation}
where $\phi=(\sqrt{2\rho_{\rm{DM}}}/m_\phi)\cos(m_\phi t+\theta)$ is the scalar DM field and $m_\phi$ denotes the mass of USDM. Eq. \ref{eq:s1} is a dimension-five operator and the coupling strength has a dimension of $1/\textrm{energy}$. The neutrino energy dispersion relation becomes
\begin{equation}
E_\nu=-g^\prime_{\alpha\beta}\dot{\phi}+|\textbf{p}|+g^\prime_{\alpha\beta}\hat{\textbf{p}}\cdot\nabla\phi+\frac{m^2}{2|\textbf{p}|}+\mathcal{O}({g^\prime}^2_{\alpha\beta}(\nabla \phi)^2).
\label{eq:s2}
\end{equation}
Hence, the effective flavour potential induced by USDM background becomes
\begin{equation}
\xi_{\alpha\beta}=-g^\prime_{\alpha\beta}\dot{\phi}+g^\prime_{\alpha\beta}\hat{\textbf{p}}\cdot \nabla\phi.
\label{eq:s3}
\end{equation}
Typically, the velocity of USDM in the Milky Way is non relativistic i.e; $v_\phi\sim 10^{-3}$, so that the first term of Eq. \ref{eq:s3} dominates. However, for our purpose, the term $\hat{\textbf{p}}\cdot\nabla \phi$ is relevant, being interested to the relative flux of neutrinos with respect to the direction associated to the $\nabla\phi$ term. In particular, we focus on the term 
\begin{equation}
\xi_{\alpha\beta}=g_{\alpha\beta}^\prime\hat{\textbf{p}}\cdot \nabla\phi\simeq g^\prime_{\alpha\beta}\sqrt{2\rho_{\rm{DM}}}\hat{\textbf{p}}\cdot \textbf{v}_\phi,
\label{eq:s4}
\end{equation}
where $\nabla \phi$ is the scalar field momentum and we use $\nabla \phi\sim m_\phi \textbf{v}_\phi \phi\sim \sqrt{2\rho_{\rm{DM}}}\textbf{v}_{\phi}$. The relative orientation of neutrino momenta with respect to $\nabla\phi$ induces an asymmetric emission of the neutrinos, contributing in such a way to the pulsar kicks. We consider time/length scale of scalar variation (coherent scalar oscillations) very large compared to the scales of the supernovae collapse processes, so that $\nabla\phi$ can be considered fixed. Proceeding in the same way as discussed in Section. \ref{sec2}, we obtain the expression for the deformation surface parameter as 
\begin{equation}
\varrho=-\frac{g^\prime_{ff} v_\phi\sqrt{2\rho_{\rm{DM}}}}{V_{\nu_f}(h^{-1}_p+h^{-1}_{V_{\nu_f}})},
\label{eq:s5}
\end{equation}
where all the quantities are measured at the resonance point. The $1\%$ of the fractional asymmetry of the neutrino momenta yields
\begin{equation}
|g^\prime_{ff}v_\phi\sqrt{2\rho_{\rm{DM}}}|=0.18V_{\nu_f}x_{\rm{res}}\lambda_\Gamma\eta.
\label{eq:s6}
\end{equation}
Using $g^\prime_{ff}\sim 1.23\times\sim 10^{-11}~\rm{eV^{-1}}$, $m_\phi\sim 1.02\times 10^{-14}~\rm{eV}$ \cite{Huang:2018cwo}, we obtain from Eq. \ref{eq:s6}
\begin{equation}
\eta=\frac{2.21\times 10^{-16}}{x_{\rm{res}}}.
\label{eq:s7}
\end{equation} 
Following the same steps as mentioned in Section \ref{sec2}, we obtain the physical solution for $x_{\rm{res}}=\mu/(\mu-1)$ and $\epsilon=[(\eta+3)\lambda_\Gamma/2\lambda_c]^{1/2}$, where all the parameters are defined in Section \ref{sec2}. To get the solution of $\mu$ we need to solve Eq. \ref{eq:32} where $\eta=2.21\times 10^{-16}$ for $\mu>1$ and $x_{\rm{res}}\sim\mathcal{O}(1)$.

In FIG. \ref{Fig1} we obtain the variation of $F(\mu)$ vs. $\mu$ for $\eta= 2.21\times 10^{-16}$. The solution of the star parameters at resonance are same as we obtain in the UVDM case. The reason is for $x_{\rm{res}}\sim\mathcal{O}(1)$, $F(\mu)$ does not change appreciably for $\eta$ given by Eq. \ref{eq:31} (UVDM) and Eq. \ref{eq:s7} (USDM). Hence, both UVDM and USDM can result pulsar kicks induced by active-sterile neutrino oscillations and the values of star parameters are same for both the cases. 


The dimension-5 operator, representing the derivative coupling between the scalar field and the neutrino field, is not arbitrary. This operator introduces a shift in the neutrino momentum, approximately proportional to $~\hat{p}.\nabla\phi$ in the neutrino dispersion relation. This shift is particularly relevant when studying asymmetric neutrino emission, which can lead to pulsar kicks.

On the other hand, the dimension-4 operator governing the interaction between the scalar field and the neutrino field is represented as $\phi\bar{\psi}\psi$. However, this interaction primarily affects the neutrino mass and does not introduce relevant terms for asymmetric neutrino emission and, consequently, pulsar kicks. 

Like ultralight scalar, ultralight pseudoscalar (axion) DM background can also explain the pulsar kicks. The axion has a derivative coupling with the neutrino current. However, there is an extra $\gamma_5$ in the neutrino current compared with Eq. \ref{eq:s1}. The analysis with axion DM will be same as USDM case since, we consider neutrinos as relativistic. 
\section{Gravitational memory signal from pulsar kicks}\label{extra}
The asymmetric neutrino emission from the core collapse supernova causes a non oscillatory permanent strain in the local spacetime metric which can be probed from the GW detectors. This phenomenon is called gravitational memory effect which is a direct consequence of Einstein's General Relativity (GR) theory. The characteristic GW amplitude produced by the neutrinos outflow is estimated by using the relation \cite{Burrows:1995bb}
\begin{equation}
h^{\rm{TT}}_{xx}=h(t)=\frac{2G}{r}\int^{t-r}_{-\infty}dt^\prime L_\nu(t^\prime)\alpha(t^\prime),
\label{m1}
\end{equation}
for transverse-traceless part of `+' mode. One can also calculate Eq. \ref{m1} for $`\times$' polarization mode. Here, $L_\nu$ denotes the neutrino luminosity and $\alpha$ denotes the anisotropy parameter. Hence, the GW strain has a strong dependence on the anisotropy parameter. We can write the upper bound of Eq. \ref{m1} as \cite{Mukhopadhyay:2021zbt}
\begin{equation}
|h(t)|\simeq \frac{2G}{r}\alpha L_\nu\mathcal{T}\lesssim 1.06\times 10^{-19}\Big(\frac{|\alpha|}{0.01}\Big)\Big(\frac{E_\textrm{tot}}{2\times 10^{53}~\rm{erg}}\Big)\Big(\frac{1~\rm{kpc}}{r}\Big),
\label{m2}
\end{equation} 
where $r$ is the distance from the source and $\mathcal{T}$ is the characteristic timescale of the neutrino emission $(E_{\rm{tot}}=L_\nu\mathcal{T})$. A geometrical definition of the quadruple anisotropy parameter is $\alpha=\frac{S_+-S_-}{S_++S_-}$, where $S_\pm$ is the area of the up/down hemisphere, whence one obtains $\alpha\simeq \frac{\Delta p}{p}=(1/18)(\varrho/r_{\rm{res}})$. In writing Eq. \ref{m2} we assume $\alpha$ is constant (typically $\alpha\simeq \mathcal{O}(0.01)-\mathcal{O}(0.1)$ \cite{Burrows:1995bb,Kotake:2009rr,Tamborra:2013laa,Walk:2019miz}). A detailed study of $\alpha$ requires to analyze the time evolution during the supernova collapse. The GW strain can also be written in frequency space for periodic and burst signals. Since, pulsars have frequencies $\mathcal{O}(\rm{kHz})$ and burst lasts for few seconds, we generally treat the source as periodic. The periodic signal can be detected in LIGO $(10-1000~\rm{Hz})$ and the strain in frequency space is given as \cite{Loveridge:2003fy}
\begin{equation}
h(f)\lesssim 10^{-24}~\rm{Hz}^{-1/2}\Big(\frac{\alpha}{0.01}\Big)\Big(\frac{10~s}{\mathcal{T}}\Big)^{1/2}\Big(\frac{1~kpc}{r}\Big)\Big(\frac{1~kHz}{f}\Big).
\label{m3}
\end{equation} 
If the signal frequency is low i.e; a few rotation of pulsar in the neutrino ejection time with frequency $0.001-0.1~\rm{Hz}$, then LISA can detect such burst signal, where the strain in frequency space depends on the anisotropy parameter as \cite{Loveridge:2003fy}
\begin{equation}
h(f)\lesssim 3\times 10^{-22}~\rm{Hz}^{-1/2}\Big(\frac{\alpha}{0.01}\Big)\Big(\frac{10~s}{\mathcal{T}}\Big)\Big(\frac{1~kpc}{r}\Big)\Big(\frac{1~Hz}{f}\Big)^{3/2}.
\label{m4}
\end{equation} 
For UVDM, the expression of $\alpha$ is 
\begin{equation}
\alpha=1.8\times 10^{-4}\sigma_1, ~~~\sigma_1=\frac{g_{ff^\prime}/m_{A^\prime}}{x_{\mathrm{res}}\eta},
\end{equation}
whereas for USDM,
\begin{equation}
\alpha=1.8\times 10^{-7}\sigma_2, ~~~\sigma_2=\frac{g_{ff^\prime}}{x_{\mathrm{res}}\eta}.
\end{equation}
The unit of $h(f)$ in frequency space is $1/\sqrt{\rm{Hz}}$ which is required to be in consistent with the definition of power spectral density, which represents the power of GW signal at different frequencies.
\begin{figure}
  \centering
\includegraphics[scale=0.6]{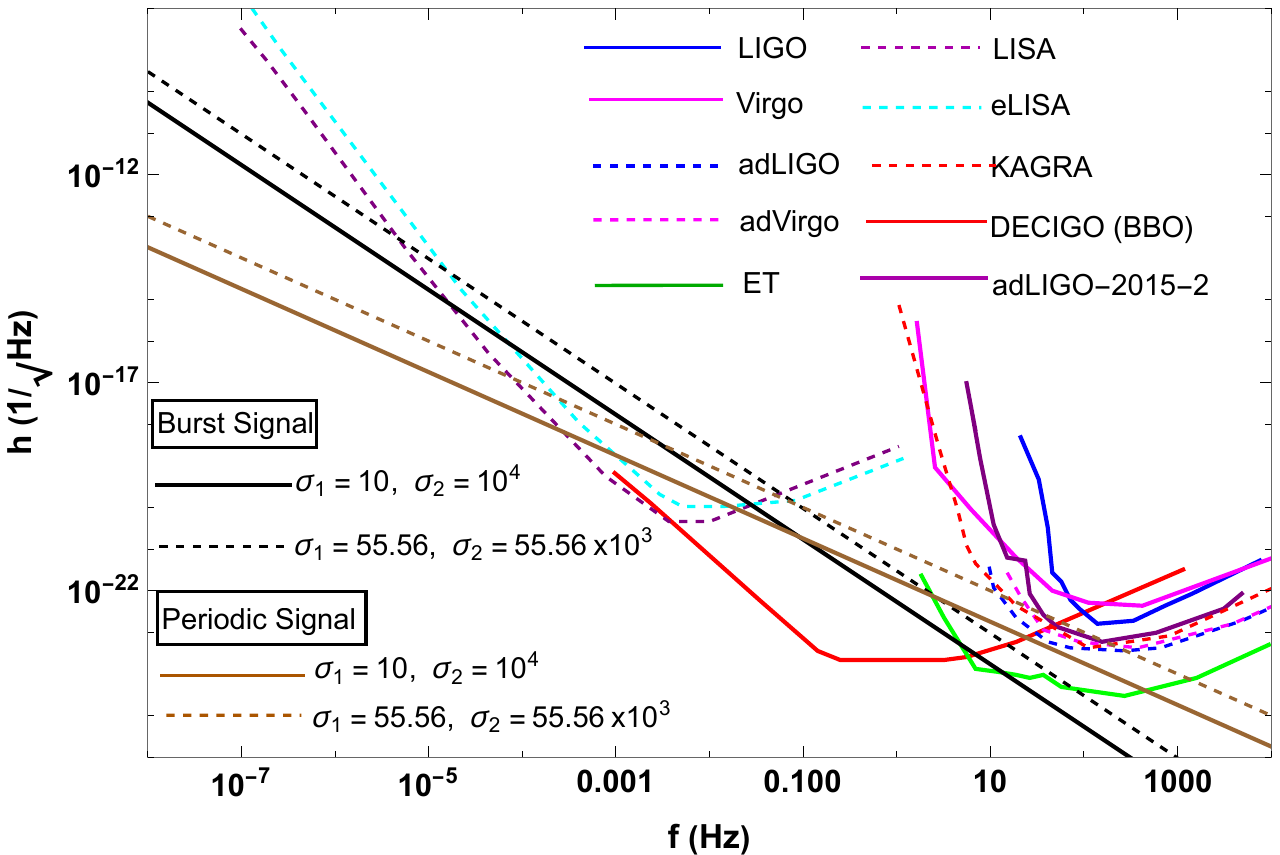}
  \caption{Variation of gravitational wave intensity with the frequency of radiation explaining a pulsar kick due to asymmetric neutrino emission caused by ULDM. The sensitivity curves of the gravitational wave detectors are shown for reference.}\label{Fig2}
\end{figure}

In FIG. \ref{Fig2} we obtain the variation of gravitational wave intensity with the frequency explaining the pulsar kick due to asymmetric neutrino emission caused by ULDM. The intensity of the GW decreases with the radiation frequency. The wave intensity increases with increasing the ULDM coupling with the neutrinos. This signal from the near supernovae with appropriate rotation frequency can be detected with the future GW detectors. The neutrino induced memory in presence of ULDM from a supernova would appear as a signal which is below the sensitivity of the present detectors (such as LIGO/Virgo) but could be observed with the second and third generation GW detectors. The rapidly rotating NSs with larger value of anisotropy parameter can be detected by the successors of present GW detectors (adLIGO, adVIRGO) whereas slowly rotating NSs can be detected by LISA. In fact, GW detectors in the deci-Hz frontiers (the region centred at $0.1~\mathrm{Hz}$) \cite{Seto:2001qf,Yagi:2011wg,TianQin:2015yph,Graham:2016plp,Ruan:2018tsw} such as DECIGO (DECi-hertz Interferometer Gravitational Wave Observatory) \cite{Seto:2001qf,Yagi:2011wg} and BBO (Big Bang Explorer) \cite{Yagi:2011wg} will reach to the sensitivity of GW strain as $10^{-24}$ and the memory signal from the pulsar kick in these detectors can be probed. The Einstein Telescope (ET) \cite{Maggiore:2019uih} will explore the universe with GWs upto cosmological distances with a sensitivity $10^{-24}-10^{-25}$ and will allow to investigate the supernova neutrinos, hence the gravitational memory effect. Also, the second and third generation GW detectors can probe smaller values of the anisotropy parameter for the asymmetric neutrino emission from a $\rm{Mpc}$ distance \cite{Mukhopadhyay:2021zbt}. Detection of such gravitational memory signal will help to constrain sterile neutrino mass, ULDM, and other beyond SM scenarios. 
\section{Relating the standard model extension parameters with the ultralight dark matter parameters from pulsar kick}\label{extra1}
We can relate the ULDM parameters with the Standard Model Extension (SME) parameters which is a direct consequence of Lorentz and CPT (Charge conjugation-Parity-Time reversal) violation \cite{Kostelecky:2003cr,Kostelecky:2004hg}. The SM is believed to be a low energy effective field theory of a more fundamental theory which confines gravity and quantum physics at the Planck scale. The SME is a theory where we add Lorentz violating operators to the SM effective field theory. These terms may or may not be CPT conserving. In the minimal SME, the general equation of motion for the neutrino fields $\nu_l$ in presence of renormalizable dimension Lorentz violating operators can be written as \cite{Kostelecky:2003cr}
\begin{equation}
(i\Gamma^{\nu}_{lm}\partial_\nu-M_{lm})\nu_m=0,
\end{equation}
where $\Gamma^{\nu}_{lm}$ and $M_{lm}$ are $4\times 4$ matrices in the spinor space and they can be written in the $\gamma$ matrices basis as \cite{Kostelecky:2003cr}
\begin{equation}
\Gamma^\nu_{lm}\equiv \gamma^\nu\delta_{lm}+c^{\mu\nu}_{lm}\gamma_\mu+d^{\mu\nu}_{lm}\gamma_5\gamma_\mu+e^{\nu}_{lm}+i f^{\nu}_{lm}\gamma_5+\frac{1}{2}g^{\lambda\mu\nu}_{lm}\sigma_{\lambda\mu},
\label{eaa1}
\end{equation}
and 
\begin{equation}
M_{lm}\equiv m_{lm}+i m_{5lm}\gamma_5+a^{\mu}_{lm}\gamma_\mu+b^\mu_{lm}\gamma_5\gamma_\mu+\frac{1}{2}H^{\mu\nu}_{lm}\sigma_{\mu\nu}.
\label{eaa2}
\end{equation}
In Eq. \ref{eaa1} and Eq. \ref{eaa2}, the mass terms $m$ and $m_5$ are CPT and Lorentz conserving. The coefficients $c$, $d$ and $H$ are Lorentz violating but CPT conserving. However, the coefficients $a$, $b$, $e$, $f$, and $g$ are both Lorentz and CPT violating. 

In SME theory, the left handed neutrino oscillation is modified by the leading order effective Hamiltonian \cite{Kostelecky:2003cr} 
\begin{equation}
h_{\mathrm{eff}}=\frac{1}{E}(a^\mu_L p_\mu-c^{\mu\nu}_L p_\mu p_\nu),
\label{ea1}
\end{equation}
where $p^{\mu}=(E,\mathbf{p})$ denotes the neutrino four momentum and $a^\mu_L=(a+b)^\mu$ and $c^{\mu\nu}_L=(c+d)^{\mu\nu}$. The Lorentz violating coefficients $a^\mu_L$ and $c^{\mu\nu}_L$ are the Hermitian matrices with mass dimension $1$ and $0$ and they correspond to CPT violating and CPT conserving operators respectively.

The expression of $\delta\mathbf{\Omega}$ for SME theory that can result the pulsar kick is given by $\delta\mathbf{\Omega}=2pc_L+a_L$ \cite{Lambiase:2005iz}. For the UVDM scenario, the SME parameter $a_L$ is related with the UVDM parameters as
\begin{equation}
a_L=g^\prime_{ff}\frac{\sqrt{2\rho_{\rm{DM}}}}{m_A^\prime}\lesssim 8.9\times 10^{-13}~\rm{eV},
\label{m5}
\end{equation}
and the $c_L$ parameter is related as 
\begin{equation}
c_L=g^\prime_{ff}\frac{\sqrt{2\rho_{\rm{DM}}}}{2p m_A^\prime }\lesssim 5.2\times 10^{-21}.
\end{equation}
For the USDM scenario, the SME parameters and the ULDM parameters are related as
\begin{equation}
a_L=g^\prime_{ff}|\nabla\phi|\lesssim 3.1\times 10^{-17}~\rm{eV},
\label{m6}
\end{equation}
and 
\begin{equation}
c_L=\frac{g^\prime_{ff}|\nabla\phi|}{2p}\lesssim 1.8\times 10^{-25}.
\end{equation}
Therefore, for the USDM case, we obtain stronger bounds on the SME parameters compared to the UVDM case. Also, the bounds on $a_L$ and $c_L$ are independent of DM mass for USDM case. Hence, the ULDM breaks Lorentz invariance and the CPT invariance to explain the pulsar kicks. This is expected because of the $\delta\mathbf{\Omega}\cdot\hat{\mathbf{p}}$ term which is responsible for the pulsar kick. The bound on $a_L$ for the USDM case is two orders of magnitude stronger than the Super-Kamiokande bound \cite{Diaz:2015dxa,Kostelecky:2008ts}. On the contrary, the bound on $c_L$ for the USDM case is two orders of magnitude weaker than the Super-Kamiokande bound \cite{Diaz:2015dxa,Kostelecky:2008ts}.
\section{Conclusions}\label{sec4}
In conclusion, we have studied the mechanism for pulsar velocities by taking into account the ultralight DM background interacting with emitted neutrinos. The effective Hamiltonian for the time evolution of neutrinos coupled with the DM field gives rise to a term of the form $\sim \delta\mathbf{\Omega}\cdot \hat{\mathbf{p}}$, where $\delta\mathbf{\Omega}=g^\prime_{ff}\frac{\sqrt{2\rho_{\rm{DM}}}}{m_{A^\prime}}$ for UVDM and $\delta\mathbf{\Omega}=g^\prime_{ff}\nabla\phi$ for USDM. The relative orientation of neutrino momenta with respect to $\delta\mathbf{\Omega}$ yields the fractional asymmetry needed for generating the pulsar kicks. Here, the ultralight DM behaves as a pseudo magnetic field. For UVDM, the ultralight vector picks a certain polarization component. For USDM, the directionality is given by $\nabla\phi$, which is basically proportional to the DM velocity vector. This generates a global asymmetry in the system for the pulsar kick. We obtain the star parameters in the resonance surface where the resonance between the active-sterile neutrinos happen in presence of ULDM background. Interestingly, the effect of the asymmetric neutrino emission induced by ULDM background can also be tested in the future GW detectors \cite{Epstein:1978dv,Burrows:1995bb,Burrows:1995ww,Loveridge:2003fy}. In addition, we relate our USDM and UVDM parameters with the SME parameter which arise in Lorentz and CPT invariance violation scenarios. 

The impacts on the neutrino oscillation experiments largely depend on the duration of one oscillation cycle of ultralight DM. The time dependent perturbation analysis should be adopted when the ultralight DM oscillates a number of cycles during the neutrino propagation. If the DM field losses its coherence within a single neutrino flight, the forward scattering effect which is coherently enhanced should vanish stochastically.  

The ultralight DM-neutrino interaction has impact on the evolution of the early universe. One can obtain bound on such couplings from CMB observables, BBN, SN1987A, and observation of blazar TXS 0506+056 which are discussed in \cite{Huang:2018cwo}. Such coupling increases the neutrino mass which increases with redshift as $(1+z)^{3/2}$ in the early universe. As a result, the neutrinos become non relativistic at the matter-radiation equality, which is in tension with current astrophysical and cosmological data. Such constraints can be avoided as discussed in \cite{Huang:2022wmz,Sen:2023uga}. 

In our manuscript, we employ the polytropic equation of state to describe the internal structure of a proto-star (we follow \cite{Barkovich:2002wh}). This mathematical model establishes a relationship between pressure, density, and temperature within the star. While this simplifies our understanding of the star formation process, wherein a gas and dust cloud contracts under gravitational forces, heats up, and eventually transforms into a proto-star, it's important to acknowledge that this model doesn't encompass all the intricate details of real proto-star evolution.

Various alternative equation of state models, such as the Murnaghan equation of state \cite{1944PNAS...30..244M} and the spherical Eddington model \cite{1982ApJ...259..311S}, have been developed to approximate proto-star evolution. However, it's worth noting that these models also come with their own inherent limitations.

More recently, hydrodynamical simulations have emerged as a more accurate and comprehensive means of elucidating the intricacies of proto-star evolution. These simulations offer a superior understanding of the processes involved in proto-star formation and evolution \cite{Nordhaus:2010ub,Scheck:2003rw,Scheck:2006rw}.

Due to the lacking of correlation between the pulsar velocity and other properties of pulsar, the statistical analysis of pulsar population neither support nor rule out any of the models in the literatures. In this paper, the mechanism responsible for pulsar kick does to rely on the pulsar velocity and magnetic field correlation. Such a result is indeed consistent with some recent analysis in which $B\rm{(magnetic~ field)}-V\rm{(velocity)}$ correlation is not needed to explain the pulsar kicks \cite{Hansen:1997zw,Toscano:1998iz,Wex:1999kr}. However, it does not mean that the Kusenko-Segr{\'e} mechanism \cite{Kusenko:1996sr,Kusenko:1998bk} will not work, because the kick depends on the magnetic field. However, this mechanism does not predict the B-V correlation as the relevant magnetic field depends on its value during the first seconds after the supernovae collapse. Observations from astrophysics only predict the surface magnetic field, millions years later. These fields are not trivially related due to its complex behaviour in the process of the cooling of neutron stars \cite{Kusenko:2004mm}.The future X-ray astronomy satellite like XRISM will be able to detect weak emission lines from trace elements and may be a new key to connect pulsar kick theories and observations. It will be also interesting to explain the pulsar kick using ultralight tensor DM \cite{Unal:2022ooa}, and in $U(1)_{L_i-L_j}$ \cite{KumarPoddar:2019ceq,KumarPoddar:2020kdz,Michimura:2020vxn} scenarios. Future improved observations and statistical analysis on pulsar velocity distribution can not only test such ultralight DM induced pulsar kicks scenarios but also it can put stringent constraints on existing models in the literature.    
\section*{Acknowledgements}
The authors would like to thank Tony Gherghetta, Satoru Katsuda, Hiroki Nagakura, Manibrata Sen, and Andrey Shkerin for useful discussions. The authors are indebted to V. Alan Kosteleck\'y for providing useful references. The authors would also like to thank the anonymous referee for the useful comments and suggestions. 
\appendix
\section{Neutrino momentum asymmetry in terms of radial deformation parameter}\label{app1}
The fractional asymmetry of the neutrino momentum can be written as (we follow \cite{Barkovich:2002wh})
\begin{equation}
\frac{\Delta p}{p}=\frac{1}{6}\frac{\int^\pi_0 \textbf{F}_S(\varphi)\cdot\delta\hat{\mathbf{\Omega}}dS}{\int^\pi_0 \textbf{F}_S(\varphi)\cdot\hat{\mathbf{n}}dS}.
\label{eq:a0}
\end{equation}
The outgoing neutrino flux $\textbf{F}_S$ which appears in the expression of neutrino momentum asymmetry fraction can be decomposed into radial and normal components as 
\begin{equation}
\mathbf{F}_S=\mathbf{F}_{\hat{n}}\hat{\mathbf{n}}+\mathbf{F}_{\hat{r}}\hat{\mathbf{r}}
\label{eq:a1}
\end{equation}
where $\mathbf{F}_{\hat{n}}$ and $\mathbf{F}_{\hat{r}}$ denote the normal and radial components of the outgoing neutrino flux to the resonance surface. The condition of local flux conservation $\nabla\cdot \textbf{F}=0$ implies that the normal and radial flux components are equal in magnitude and they are both equal to the half of the diffusive outgoing flux $\mathbf{F}_\nu$. Hence, we can write
\begin{eqnarray}
F_{\hat{n}}(\varphi)&=&F_{\hat{r}}(\varphi)\nonumber\\
&=& \frac{1}{2}F_{\nu}(r_{\rm{res}}+\varrho\cos\varphi)\nonumber\\
&=& \frac{1}{2}F_{\nu}(r_{\rm{res}})(1+h^{-1}_F\varrho\cos\varphi),
\label{eq:a2}
\end{eqnarray}
where $h^{-1}_F=\frac{1}{F}\frac{dF}{dr}\Big|_{r_{\rm{res}}}$. Also, the elemental area $dS$ is related with the elemental area of the resonance surface as 
\begin{equation}
dS\simeq\Big(1+2\frac{\varrho}{r_{\rm{res}}}\cos\varphi\Big)dS_r,
\label{eq:a3}
\end{equation}
and 
\begin{equation}
\hat{\mathbf{r}}\cdot \delta\hat{\mathbf{\Omega}}=\cos\varphi, \hspace{0.5cm} \hat{\mathbf{n}}\cdot\delta\hat{\mathbf{\Omega}}\simeq\cos\varphi-\frac{\varrho}{r_{\rm{res}}}\sin^2\varphi.
\label{eq:a4}
\end{equation}
Using Eq. \ref{eq:a0}, Eq. \ref{eq:a1}, Eq. \ref{eq:a2}, Eq. \ref{eq:a3}, and Eq. \ref{eq:a4} we can write the neutrino momentum asymmetry as
\begin{equation}
\frac{\Delta p}{p}=\frac{1}{36}h^{-1}_F\varrho,
\label{eq:a5}
\end{equation}
where only the normal component of the neutrino flux contributes to the numerator of the integral in Eq. \ref{eq:a0}. Hence, the non zero value of the gradient of the outgoing neutrino flux $(h^{-1}_F\neq 0)$ contributes to the pulsar kick. Using $h^{-1}_F\sim -2r^{-1}_{\rm{res}}$, we obtain \cite{Barkovich:2002wh}
\begin{equation}
\frac{\Delta p}{p}\simeq-\frac{1}{18}\frac{\varrho}{r_{\rm{res}}}.
\label{eq:a6}
\end{equation}
\section{The Polytropic Model}\label{app2}
We recall the main topics of the polytropic model which describes the inner core of a proto-star (we follow \cite{Barkovich:2002wh}). The isotropic neutrinosphere is described by the following equations
\begin{eqnarray}\label{eq:b1}
  \frac{dP(r)}{dr}&=&-\rho(r)\, \frac{GM(r)}{r^2} \,\, (\text{ hydrodynamical equilibrium}) \\
   F(r) &=& -\frac{1}{36} \frac{1}{\kappa \rho(r)}\frac{dT^2}{dr} \,\, (\text{energy transport}) \label{eq:b2} \\
   F(r) &=& \frac{L_c}{4\pi r^2} \,\, (\text{flux conservation}), \label{eq:b3}
\end{eqnarray}
where $M(r)=4\pi\int_0^r dr^\prime {r^\prime}^2\rho_T(r^\prime)$ ($\rho_T$ is the total density of matter) and $L_c$ is the luminosity of the proto-star. The equation of state of a polytropic gas filled with relativistic nucleons characterizing the proto-star is given as \cite{Barkovich:2002wh}
\begin{equation}
P(r)=K\rho^\Gamma\,,\quad K=\frac{T_c}{m_n\rho_c^{1/3}},
\label{eq:b4}
\end{equation}
where $\Gamma$, $T_c$ and $\rho_c$ are the adiabatic index, temperature and matter density of the core respectively. Using Eq. \ref{eq:b1} and Eq. \ref{eq:b4} we obtain 
\begin{equation}
\frac{d\rho^{\Gamma-1}}{dr}=-\frac{\lambda_\Gamma r_c\rho_c^{\Gamma-1}}{M_c}\frac{M(r)}{r^2},
\label{eq:b5}
\end{equation}
where $\rho_c$, $M_c$ and $r_c$ are the density, mass and radius of the core respectively. Here, $\lambda_\Gamma=(GM_c/r_c\rho_c^{\Gamma-1})(\Gamma-1)/K\Gamma$. The density $\rho$ can be expressed, with extremely good approximation as \cite{Barkovich:2002wh}
\begin{equation}
\rho^{\Gamma-1}(r)=\rho^{\Gamma-1}_c\Big[\lambda_\Gamma\Big(\frac{r_c}{r}-1\Big)m(r)+1\Big],
\label{eq:b6}
\end{equation}
where $m(r)=\mu+(1-\mu)r_c/r$ and hence, $m(r_c)=1$. The condition $\rho(R_s)=0$ ($R_s$ is the radius of the star) allows to determine the parameter $\mu$, i.e; Eq. \ref{eq:19}. Moreover, Eq. \ref{eq:b6} can be easily recast in the form Eq. \ref{eq:17}. The temperature profile follows from Eq. \ref{eq:b2} and Eq. \ref{eq:b3} as
\begin{equation}
\frac{dT^2}{dr}=-\frac{9\kappa L_c}{\pi r^2}\rho,
\label{eq:b7}
\end{equation}
with $\rho$ given by Eq. \ref{eq:17}. For $\Gamma=4/3$, the solution of Eq. \ref{eq:b7} becomes \cite{Barkovich:2002wh}
\begin{equation}
T(r)=T_c\sqrt{2\lambda_c\Big[\chi(r_c/r)-\chi(1)+1\Big]},
\label{eq:b8}
\end{equation}
with $\chi(r_c/r)$ defined in Eq. \ref{eq:22}.
\bibliographystyle{utphys}
\bibliography{bira}
\end{document}